\documentclass[apjl]{emulateapj}

\usepackage{xcolor}
\usepackage{natbib}
\usepackage{hyperref}

\newcommand{\GE}[0]{{{\it G}-E}}
\newcommand{\eagle}[0]{{\sc eagle}}

\newcommand{\MsunInline}[0]{{M$_\odot$}}

\shorttitle{A Gaia-Enceladus analog in the EAGLE simulation}
\shortauthors{Bignone et al.}

\begin{document}

\title{A {\it Gaia}-Enceladus Analog in the EAGLE Simulation: \\ Insights
  into the Early Evolution of the Milky Way}


\author{Lucas A. Bignone\altaffilmark{1,2}, Amina Helmi\altaffilmark{3}, Patricia B. Tissera\altaffilmark{2}}

\begin{abstract}
  We identify a simulated Milky Way analog in the \eagle{} suite of
  cosmological hydrodynamical simulations. This galaxy not only shares
  similar global properties as the Milky Way, but was specifically
  selected because its merger history resembles that currently known for
  the Milky Way. In particular we find that this Milky Way analog has
  experienced its last significant merger (with a stellar mass ratio $\sim
  0.2$) at $z\sim 1.2$. We show that this merger affected both the dynamical
  properties of the stars present at the time, contributing to the
  formation of a thick disk, and also leading to a significant increase
  in the star formation rate of the host. This object is thus particularly
  suitable for understanding the early evolutionary history of the Milky Way. It
  is also an ideal candidate for re-simulation with much higher resolution as
  this would allow addressing a plethora of interesting questions such as, for
  example, the specific distribution of dark matter near the Sun.
\end{abstract}

\keywords{Stellar kinematics -- Milky Way formation -- Galaxy mergers}

\altaffiltext{1}{Corresponding author; {\sf{email: l.bignone@uandresbello.edu}}}
\altaffiltext{2}{Departamento de Ciencias Fisicas, Universidad Andres Bello, 700 Fernandez Concha, Las Condes, Santiago, Chile}
\altaffiltext{3}{Kapteyn Astronomical Institute, University of Groningen, 
 P.O. Box 800, 9700 AV Groningen, The Netherlands}

\section{Introduction}
\label{sec:Introduction}

Very significant progress in our understanding of the assembly history of the
Milky Way (MW) has recently been made. This was enabled by the {\it Gaia}
mission second data release \citep{collaboration_gaia_2018} in conjunction with
large spectroscopic surveys such as the APO Galactic Evolution Experiment
\citep[APOGEE][]{abolfathi_fourteenth_2018}, Galactic Archaeology with HERMES
\citep[GALAH][]{buder_galah_2018}, the RAdial Velocity Experiment
\citep[RAVE][]{kunder_radial_2017} and the Large Sky Area Multi-Object Fibre
Spectro- scopic Telescope \citep[LAMOST][]{zhao_lamost_2012,cui_large_2012}. A
striking result was the diskovery \citep{helmi_merger_2018} that a large
fraction of the inner halo was made up of debris from a single galaxy named {\it
Gaia}-Enceladus (\GE{}, hereafter), which is sometimes also referred to as as
the {\it Gaia} sausage because of the signature in velocity space
\citep{belokurov_co-formation_2018}. This was the last significant merger that
the Galaxy experienced and was estimated to have taken place $\sim10$ Gyr ago.
It is likely that this event also had an impact on the disk present at the time,
but the exact details of this process are not yet known
\citep{gallart_uncovering_2019}. 

The key objective of this Letter is to identify, in a state-of-the-art
cosmological simulation, an MW analog with a history of assembly that better
resembles our current knowledge of the MW, including a merger event of similar
characteristics as that found by \citet{helmi_merger_2018}. This would
particularly aid in understanding its effects. Such a system would also be an
excellent candidate for re-simulation \citep[via the zoom-in
technique,][]{katz_formation_1994,navarro_simulations_1994}, and allow for a
more direct comparison of observations and models. Furthermore, such a
re-simulation could be especially useful for studying the distribution of dark
matter near the Sun (a critical input of dark matter direct detection
experiments; \citep{herzog-arbeitman_empirical_2018, necib_under_2018}), as well
as give new insights into the detailed link between the formation of various
Galactic components, and possibly also other peculiarities about the MW, such as
the $\alpha$ patterns and stellar age of the inner halo
\citep{fernandez-alvar_assembly_2019, carollo_origin_2018} and the thick disk
\citep{carollo_evidence_2019}.

In this {\it Letter} we analyze in detail a simulated galaxy extracted from the
EAGLE suit \citep{schaye_eagle_2015,crain_eagle_2015}, a series of cosmological
hydrodynamic simulations. This simulated galaxy has an assembly history similar
to that currently known for the MW. We study in detail the properties of
this MW-analog, and in particular focus on the effects on the different
galactic components present at the time of the merger with a \GE{}-like
system. 

\begin{figure*}
    \centering
        \includegraphics[width=\textwidth]{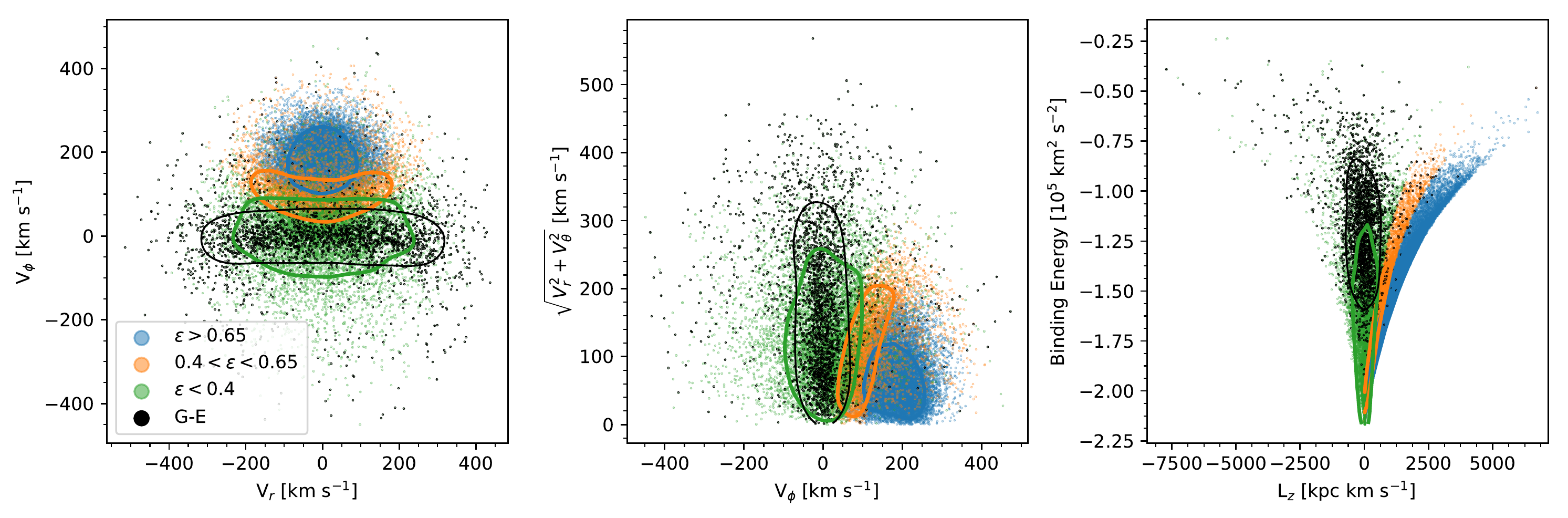}
        \caption{Left panel: velocity distribution of stars in spherical
          coordinates, radial $v_r$ and azimuthal $v_\phi$ for stellar
          particles in an \eagle{} galaxy that had a \GE{}
          type merger event. Black points represent the
          accreted stars from the \GE{} analog. Other
          colors represent stars in the circularity ranges displayed
          in the legend. Contours envelop 68\% of particles in
          each subsample. Middle panel: Toomre diagram of stellar
          particles. Note that the accreted stars have slightly
          retrograde mean motions. Right pnale: distribution of the particle
          binding energy vs. angular momentum, where the accreted stars
          have lower binding energies and slightly retrograde angular
          momentum, which allows for easier identification (especially
          for $E > -1.25\times 10^5$~km$^2$ s$^-{2}$). Due to the limited
          resolution of the simulation, we plot in the three panels the velocities of all star
          particles and not only those in the solar vicinity}
    \label{fig:merger_kinematics}
\end{figure*}

\section{\GE{} analog identification}
\label{sec:analog_identification}

The EAGLE project \citep[see][for details]{schaye_eagle_2015} has been shown to
produce a realistic population of galaxies reproducing a broad range of observed
galaxy properties and scaling relations. Here we concentrate on the largest
EAGLE simulation (Ref-L100N1504\footnote{Data extracted from the \eagle{} public
data release \citep{mcalpine_eagle_2016}}) which has a comoving cubic volume of
100 Mpc in linear extent. The mass resolution of dark matter is
$\sim9.7\times10^6$~\MsunInline{} and the initial mass of baryonic particles is
$\sim1.81\times10^6$ \MsunInline{}. While the resolution of this particular
simulation is lower than others available (the gravitational softening length is
limited to a maximum physical size of 0.70 kpc), its large volume provides
numerous MW-type galaxies with a wide range of merger histories.

As a first step in our identification of a suitable MW analog with a \GE{}-like
merger, we select galaxies with virial masses in the expected range for the MW,
M$_{200}$ = [1, 1.5] $\times 10^{12}$ \MsunInline{}
\citep[e.g.][]{posti_mass_2019,watkins_evidence_2019}. We also impose that the
stellar mass inside a 30 kpc radius be greater than $10^{10}$ \MsunInline{} and
that the current star formation rate (SFR) be in the range
0.1-3~\MsunInline{}~yr$^{-1}$. These simulated galaxies are then decomposed
dynamically into a spheroid and a disk component \citep{tissera_chemical_2012},
where the disk is defined by star particles with circularity $\epsilon=L_z/L_{z,
\rm{max}}(E) > 0.5$, with $L_{z, \rm{max}}(E)$ being the maximum angular
momentum along the main axis of rotation, over all particles at a given binding
energy, $E$. A stellar disk-to-total mass ratio criterion (D/T) $> 0.4$, is then
imposed to ensure that our sample comprises galaxies with significant disks.

Finally, we diskard galaxies that experienced a merger event with a
stellar mass ratio $> 0.15$ at any redshift $z<1$. This constraint comes
from the fact that the Sagittarius dwarf galaxy, with its stellar mass lower
than $5 \times 10^8$\MsunInline{}, is likely the biggest merger since $z \sim 1$
\citep{ibata_dwarf_1994}.  The final sample comprises 101 MW-like galaxies
with diverse merger histories for $z\ge1$.

A characteristic of the \GE{} stars is that they are on very eccentric orbits
\citep{belokurov_co-formation_2018,myeong_milky_2018}. Therefore, we look for
similar signs of radial anisotropy in the galaxies of our sample. To identify
halo stars, we select stellar particles with $\epsilon < 0.4$ and vertical
distance from the disk plane $|Z| > 5$ kpc (this last restriction is to diminish
the possible contamination from disk or bulge stars). We then characterize the
shape of the stellar halo's velocity ellipsoid by the anisotropy parameter:
$\beta = 1 - \frac{\sigma_\theta^2 + \sigma_\phi^2}{2\sigma_r^2}$, where
$\sigma_i$ are the velocity dispersions in spherical coordinates. We consider in
the computation of $\beta$ only stellar particles within 5 kpc $< R < 12$ kpc,
where $R$ is the radial distance of the particles in the plane of the disk.

The final sample of (101) MW-like galaxies has a median $\beta$ of 0.46. If we
assume that a halo dominated by the kinematic anisotropic component  has $\beta$
more than $2\sigma$ higher than the median (i.e. $\beta > 0.68$), five MW-like
galaxies are left. We analyzed their assembly histories, searching for the
accreted satellite galaxies that contributed more significantly to  the
anisotropic distribution and had a stellar mass ratio that is comparable to the
value estimated for \GE{} by \citet{helmi_merger_2018} and  occurred around the
estimated time by \citet{helmi_merger_2018,hawkins_relative_2014}. Only one of
the selected halos satisfies these constraints and also has a value of $\beta
\sim 0.73$, which is comparable to that found for the dynamical structure
commonly referred to as the {\it Gaia} Sausage \citep{belokurov_co-formation_2018,
fattahi_origin_2019}. While variations in the selection criteria could result in
some additional candidates, this kind of events are expected to be rare. For
example, \citet{mackereth_origin_2019} studied the origin of highly eccentric
accreted stars in a smaller volume of the EAGLE simulation suit and found that
only $\sim14$\% of MW-mass galaxies had an accretion profiles resembling the GE
observations, using less stringent criteria.

\begin{deluxetable}{lcccc} 
  \centering
  \tablecaption{Properties of the MW and the \GE{} analogs.}
  \tablehead{
    \colhead{Property} & \multicolumn{2}{c}{MW} & \multicolumn{1}{c}{\GE{}} & \colhead{Unit} \\
    & \colhead{($z=0$)} & \colhead{($z=1.7$)} & \colhead{($z=1.7$)} 
  }
  \startdata
  Virial mass, M$_{200}$ & $1.2\times10^{12}$ & $8.3\times10^{11}$ & $(1.5\times10^{11})$\tablenotemark{a} &\MsunInline{}\\
  Virial radius, R$_{200}$ & $226.4$ & 113.0 & $(53.2)$\tablenotemark{a} & kpc\\
  Gas mass, M$_\textnormal{gas}$ & $6.8\times10^{9}$ & $1.9\times10^{10}$ & $3.6\times10^{9}$ & \MsunInline{}\\
  Stellar mass, M$_*$ & $4\times10^{10}$ & $1.5\times10^{10}$ & $3.1\times10^{9}$ & \MsunInline{}\\
  M$_{*, \epsilon>0.65}$ & $2.2\times10^{10}$ & $4\times10^{9}$ & $4.4\times10^{8}$ & \MsunInline{}\\
  M$_{*, 0.4<\epsilon<0.65}$ & $6.4\times10^{9}$ & $4.1\times10^{9}$ & $5\times10^{8}$ & \MsunInline{}\\
  M$_{*, \epsilon<0.4}$ & $1.3\times10^{10}$ & $7.2\times10^{9}$ & $2.1\times10^{9}$ & \MsunInline{}\\
  $D/T$ & 0.64 & 0.42 & 0.23 & \\
  \enddata
  \tablenotetext{a}{Property computed at $z=2$, before the start of the merger.}
\label{tab1}
\end{deluxetable}

\begin{figure*}
  \centering
      \includegraphics[width=\textwidth]{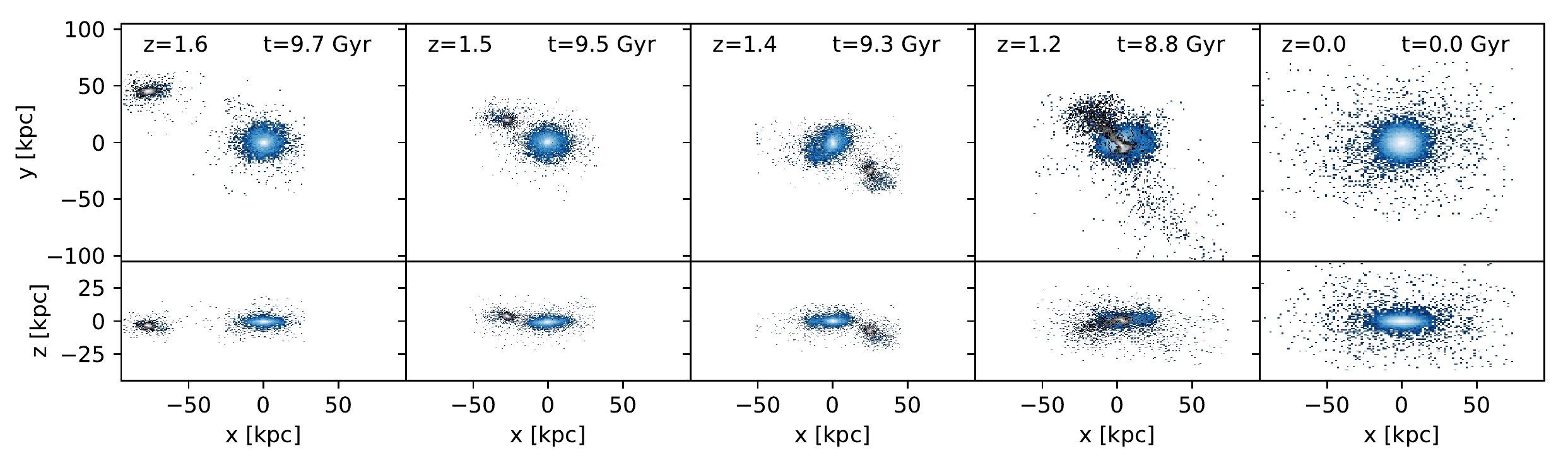}
      \caption{Face-on and edge-on views of the spatial distribution of stars in
      the MW (blue points) and \GE{} (black points) analogs during the merger.
      The reference frame has been centered on the center of potential of the
      host disk and has also been rotated to align the angular momentum of the
      main disk with the {\it z-axis} at each timestep. The last panel shows the
      final configuration of the MW analog at $z=0$}

  \label{fig:snapshots}
\end{figure*}

At the time of the merger, the selected MW analog had a stellar mass
M$_*\sim1.5\times10^{10}$ \MsunInline{} and a distinct disk morphology
($D/T\sim0.42$), while the \GE{} analog had M$_*\sim3.1\times10^{9}$
\MsunInline{} and an irregular morphology ($D/T\sim0.23$). Table \ref{tab1}
lists the general properties of the MW analog at $z=0$ and again at $z=1.7$, a
time just before the \GE{} analog crossed its virial radius. The properties
for the \GE{} analog are also listed in Table \ref{tab1} for $z=1.7$.

Fig.~\ref{fig:merger_kinematics} shows the dynamical properties of all stars
from our MW analog at $z=0$ with galactocentric distances $r_{\text{gc}} < 30$
kpc. We have separated the stars into three groups according to their
circularity: $\epsilon > 0.65$, $0.4 < \epsilon < 0.65$ and $\epsilon < 0.4$.
These groups can be roughly considered as constituting, respectively, a
``fiducial'' thin disk (blue), a thick disk (orange) and a spheroid (i.e. bulge
and stellar halo, green). We refer to these components as ``fiducial'' because
the numerical resolution of the simulation prevents us from obtaining a detailed
definition of thin and thick disks. However, we can group stellar particles
according to the circularity as an indicator of degree of rotational support. We
plot separately the stars accreted from the \GE{} analog (black points). The
first panel of Fig.~\ref{fig:merger_kinematics} shows their $v_r$--$v_\phi$
velocity distribution and reveals that the accreted stars have a sausage-like
distribution in velocity space due to their highly elongated orbits.

The middle panel of Fig.~\ref{fig:merger_kinematics} shows the Toomre diagram
again for all stellar particles in our MW analog. In agreement with the results
reported by \citet{koppelman_one_2018} and \citet{helmi_merger_2018} for halo
stars near the Sun, the accreted stellar particles present slightly mean
retrograde rotation. The right panel of Fig.~\ref{fig:merger_kinematics} shows
the stellar particles binding energies versus $L_z$, where we see that the
accreted particles belong almost completely to the $\epsilon < 0.4$ spheroid
component. As indicated by the contours in this figure, the accreted stars are
less gravitationally bounded than the majority of the spheroid stars. Together
with their mean retrograde motion, this makes them easy to distinguish from
particles belonging to the other components. For example, a clear gap can be
observed between the accreted stars and the particles with low binding energy
belonging to the thick disk, which is similar to what has been observed in the
MW's solar vicinity \citep{koppelman_one_2018,helmi_merger_2018}.

The qualitative similarities between the kinematics of halo stars near the Sun
and the merger debris in our simulated MW-analog warrants a more detailed
analysis. In the following section, we thus proceed to characterize the merger
and determine the effects that it had on the evolution of the different
components of the final galaxy.

\section{Results}
\label{sec:Results}

\begin{figure*}
    \centering
        \includegraphics[width=\textwidth]{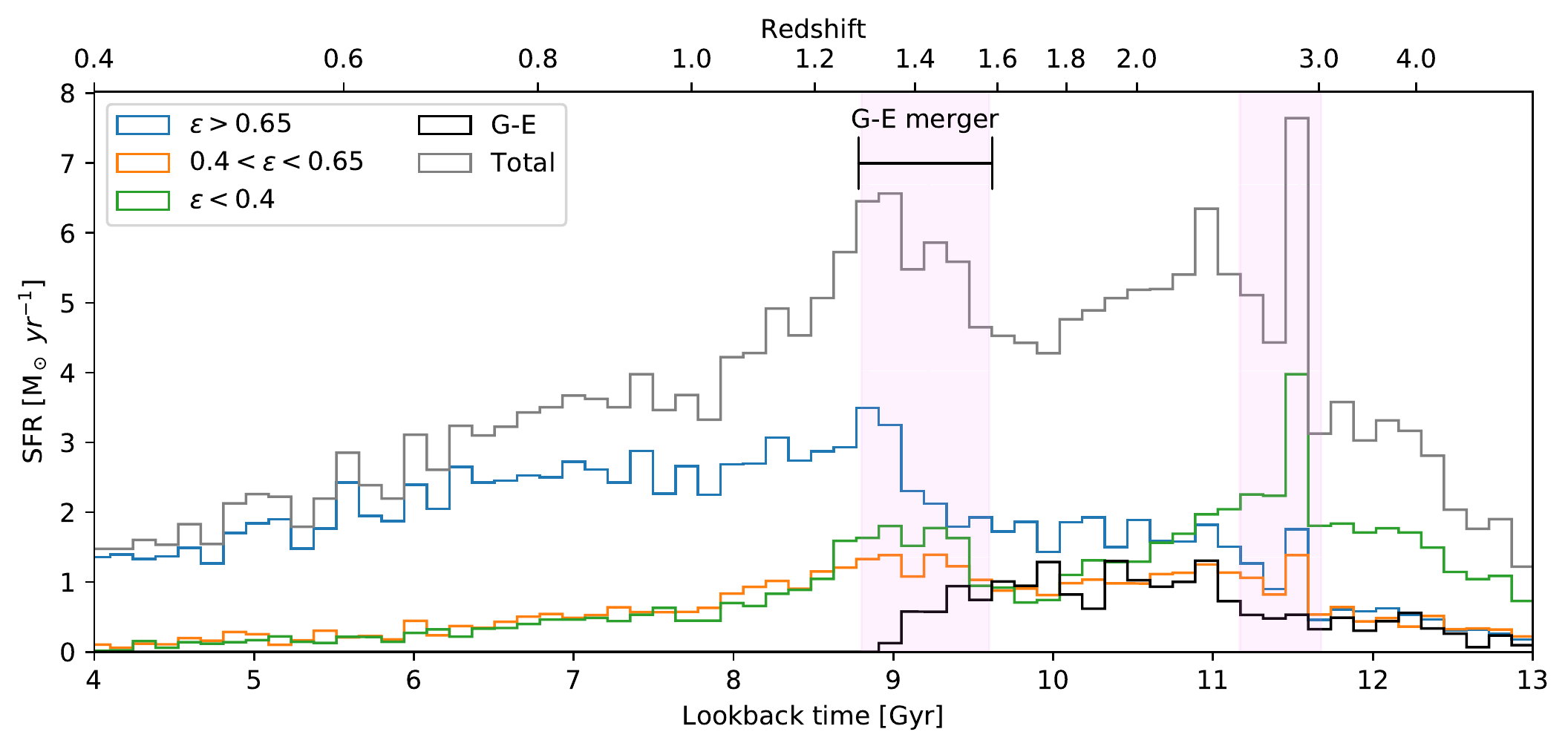}
        \caption{Star formation history of the host galaxy, diskriminating stars
          in three circularity ranges at $z=0$ as in
          Fig.~\ref{fig:merger_kinematics}. In black, we show the star formation
          history of stellar particles accreted from the satellite. The gray
          line shows the star formation history of all stellar particles. A
          starburst coinciding with the time of the merger (shaded interval) can
          be seen for all stellar particles. The shaded region on the left
          corresponds to a previous major merger with stellar mass ratio 0.8.}
    \label{fig:sf_history_filtered}
\end{figure*}

To analyze the merger between the simulated \GE{} and MW galaxies, we have used
data from the \eagle{} snipshots, which have a temporal resolution of $\sim90$
Myr, which is significantly higher than that of the standard snapshots. At each
timestep, we have identified the planes of rotation of the disks belonging to
the main progenitor of the MW and \GE{} analogs. In Figure \ref{fig:snapshots},
we show face-on and edge-on views of different stages of the merger.

\subsection{Orbital properties}

We determine the orbital properties of the encounter at the time just before the
satellite entered the virial radius of the main galaxy. The orbit of the
satellite is almost coplanar with respect to the plane of the disk of the host
with an inclination $i=6.6^{\circ}$. The encounter is retrograde and the
satellite's spin is counter-rotating with respect to that of the main galaxy.
Together, these orbital properties explain why the merger debris has retrograde
orbits. Based on isolated simulations of minor mergers of (prograde rotating)
disks \citep{villalobos_simulations_2008}, \citet{helmi_merger_2018} found that
the mean retrograde motion measured for \GE{} debris were best matched in the
case of retrograde encounters with inclinations between $30^\circ$ and
$60^\circ$. Similarly, we do find that the encounter is ``net'' retrograde,
although the inclination of our merger is lower than theirs.

In our simulation, the satellite reaches $\sim 4.9$ kpc at closest approach in
its first passage, which is then followed by a maximal separation of $\sim 40$
kpc. The satellite has completely merged after a second passage. The dynamical
effects of the merger thus lasts about 0.8 Gyr between lookback times 8.8 Gyr
($z=1.24$) and 9.6 Gyr ($z=1.53$).

\subsection{Star formation history}

In Figure \ref{fig:sf_history_filtered} we show the star formation history of
the three fiducial components identified in the MW analog at $z=0$. We have
explicitly removed from the computation the stellar particles accreted from the
\GE{} analog, which we show separately (in black).
Fig.~\ref{fig:sf_history_filtered} shows an overall increase in the SFR around
the time of the merger (shaded region), suggesting that the \GE{} analog induced
star formation in the MW progenitor during their interaction. The spheroid and
the fiducial thick disk also show an increase of the SF activity during this
period. Star formation in these components continues after the merger, although
at a declining rate until it is quenched almost completely $\sim4$ Gyr ago. The
effect on the fiducial thin disk, on the other hand, appears only toward the
end of the merger. The gas fraction of the \GE{} is $f_\textnormal{gas}\sim0.6$,
which amounts to a gas mass of $\sim4.5\times10^{9}$ \MsunInline{} out of which
$\sim2.9\times10^{9}$ \MsunInline{} are converted into stars in the thin disk,
representing $\sim13$ \% of the stellar mass of the thin disk at $z=0$. \GE{}
thus contributes a significant amount to the subsequent growth of the MW, but
does not accounts for all of it.

Figure \ref{fig:sf_history_filtered} shows the existence of other starbursts,
particularly noticeable is that at $z \sim 2.8$ and which is clearly evident in
the spheroid component. This corresponds to another important merger with a
stellar mass ratio of 0.8 that occurred around that time in this MW analog.

\begin{figure*}
    \centering
        \includegraphics[width=\textwidth]{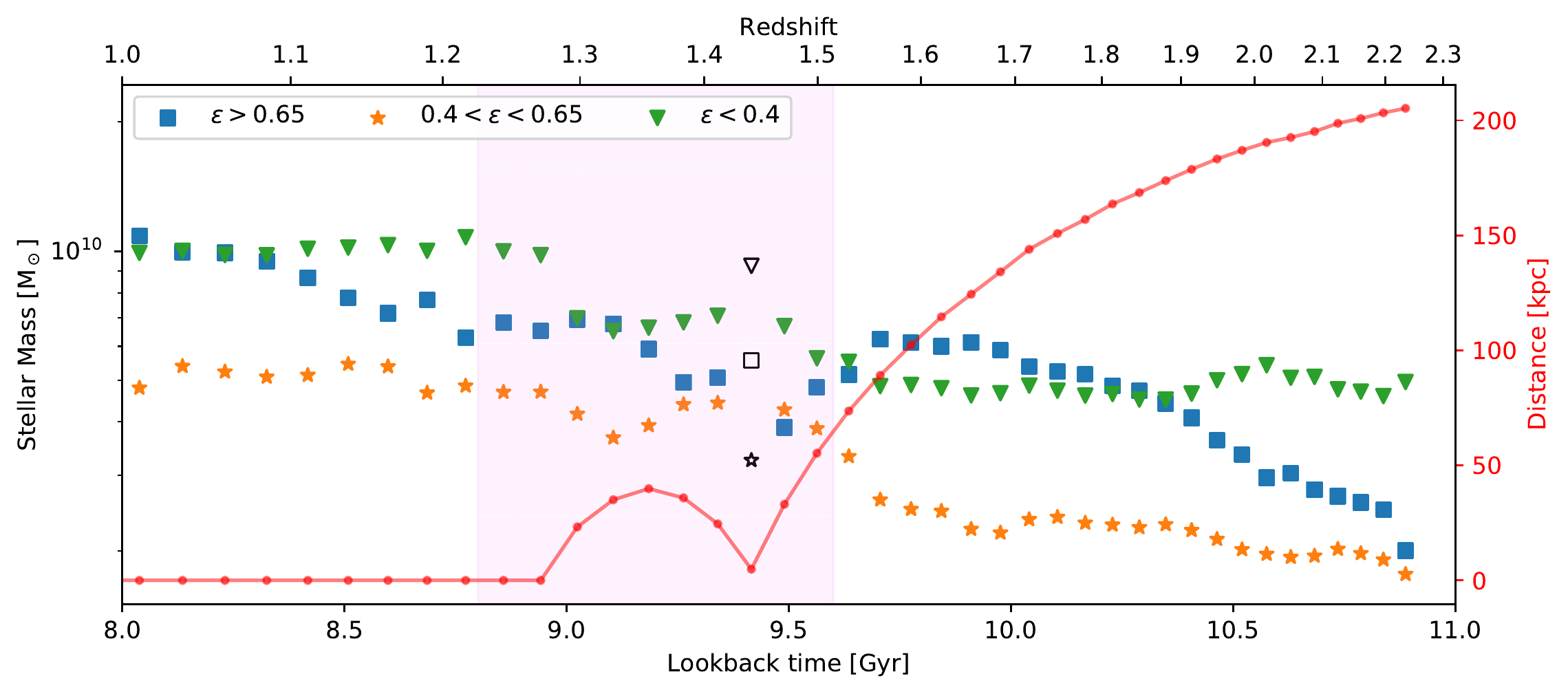}
        \caption{Evolution of the stellar mass content with lookback time in the
          host galaxy for stellar particles in three circularity ranges. The
          stellar mass in particles with $\epsilon < 0.4$ presents a marked
          increase due to the accretion of stars from the satellite. The stellar
          mass with intermediate circularities ($0.4 < \epsilon < 0.65$) also
          presents an increase just before and during the merger that can be
          attributed to a thickening of the disk triggered by the merger (see
          the text for details). The red line represents the distance between
          the centers of mass of the satellite and its host galaxy. The shaded
          region marks the lookback times between 8.8 and 9.6 Gyr, when the
          merger is expected to have its most significant dynamical effects. At
          the time of the closest approach between \GE{} and the MW, the
          assignment of particles between the host and the satellite becomes
          unreliable. We mark with empty symbols the points for which
          contamination from \GE{} particles is expected and exclude them from
          our analysis. }
    \label{fig:components_mass_evolution}
\end{figure*}

\subsection{Evolution of the structural components}

We now zoom-in around the time of the merger with the \GE{} analog to quantify
the effect that it had on the growth of the fiducial thin disk, thick disk and
spheroid components. To do so, we identify the plane of rotation of the MW
analog main progenitor at each timestep and sort the stars in each progenitor
galaxy into three groups according to their circularity, in the same way that we
did in section \ref{sec:analog_identification} for $z=0$.

In Figure \ref{fig:components_mass_evolution} we show the stellar mass
present in each component  as function of lookback
time. We also show in the figure the galactocentric distance between
the centers of mass of the MW and \GE{} analogs.  We find that the
mass of the spheroid component (green line) grows by 122\% between the
start and the end of the \GE{} merger. The mass of the fiducial thick
disk (orange line) also grows considerably, by 84\% during the merger. On
the other hand, the fiducial thin disk (blue) starts growing more prominently
during the last stage of the merger, as we had already seen in
Fig.~\ref{fig:sf_history_filtered}.

Part of the growth of the spheroid component during this time interval can
be directly attributed to captured stars from the satellite, as this deposits
96\% of its stars in this component. This in turn represent 19\% of the $\epsilon < 0.4$
stars at $z=0$. Additional growth of this component comes from in situ star
formation, most of which we can attribute to the starburst shown in Figure
\ref{fig:sf_history_filtered}. 

At disk heights $|Z| > 5$ kpc and $r_\textnormal{gc} < 30$ kpc, 38\% of the
stars came from the \GE{} analog, which supports the idea that most of the MW
inner halo formed from a merger that is similar to the one presented here.

The increase of stellar mass in the fiducial thick disk, however, is not due to
direct accretion of stellar particles (only 3.5\% of the satellite's stellar
mass ends up in this component, representing 1.4\% of the mass of this component
at $z=0$, but is instead due to the decrease of the circularity of stars that
originally belonged to the thin disk. This could be interpreted as a result of
the dynamical heating induced by the merger. We also noticed that this transfer
of particles between components mostly occurs just before and during the first
passage of the satellite. This confirms the suggestion by
\citet{helmi_merger_2018} and \citet{haywood_disguise_2018} that the merger with
Gaia-Enceladus could have played a significant role in the establishment of the
MW's thick disk as a distinct kinematical structure; see also
\citet{carollo_evidence_2019,gallart_uncovering_2019}. Despite the interaction
and being partially heated, the primordial thin disk is not destroyed and
remains largely in place during and after the merger.

Finally, we note that less than 1\% of the stars belonging originally to the
\GE{} analog have $\epsilon > 0.65$ at $z=0$ and they only represent 0.06\% of
the final fiducial thin disk, which is in agreement with previous numerical
results \citep{tissera_chemical_2012,gomez_lessons_2017}.

\section{diskussion and Conclusions}

We have analyzed a merger between a MW-like galaxy and a {\it Gaia}-Enceladus
analog extracted from the large cosmological box of the \eagle{} simulation. We
arrived at our object of study by first selecting a sample of galaxies that
closely resemble the MW in terms of its present-day properties (stellar mass,
virial mass, SFR, D/T ratio, and $z<1$ assembly history), and second, by
determining which galaxy presented the largest radial velocity anisotropy
feature in the halo stars that could be similar to the so-called {\it Gaia}
sausage. Curiously, we have found that the merger debris in our simulations is
on a slightly retrograde orbits as also found for {\it Gaia}-Enceladus. The mean
retrograde motions are in our case the result of a nearly coplanar merger
between counter-rotating disks.

In the selected MW galaxy, the merger with the \GE{} analog took place between
lookback times $\sim8.8$ Gyr and $\sim9.6$ Gyr and had important effects on the
evolution of the host galaxy. Our analysis shows how this merger not only
induced starbursts in the early disk, as expected based on previous
observational \citep[e.g.][]{larson_star_1978,lambas_galaxy_2003-1} and
numerical
\citep[e.g.][]{mihos_gasdynamics_1996-1,barnes_shock-induced_2004,sillero_evolution_2017}
results, but additionally led to the formation of co-eval stellar populations
that are part of the spheroid and thick disk of the MW-analog at $z=0$.

Furthermore, as a result of the interaction with \GE{}, stars in the early thin
(and thick disk) were dynamically heated, forming the present-day thick disk. We
found that most of the debris of the \GE{} merger is part of the spheroid at the
present time. A large fraction of its debris in our simulation is deposited at
large heights, corresponding to the stellar halo.

The simulation is therefore in agreement with a scenario where the MW
experienced a significant merger early in its history, evidence of which can be
found in the present-day kinematic properties of halo stars. Furthermore, there
is rough agreement between the stellar mass of the \GE{} analog
($\sim3.1\times10^{9}$ \MsunInline{}) and that predicted for the event
$\sim5\times10^9$ \MsunInline{}
\citep{belokurov_co-formation_2018,helmi_merger_2018,fernandez-alvar_assembly_2019,mackereth_origin_2019,vincenzo_fall_2019}.

In terms of the timing of the event, the observed ages between 10 and 13 Gyr of
\GE{} stars as inferred by isochrones
\citep{helmi_merger_2018,hawkins_relative_2014} is compatible with the stellar
ages of the \GE{} analog, for which a median stellar age of $10.8$ Gyr is found.
This is also in agreement with results by \citet{mackereth_origin_2019}, who
found that accreted stars in \eagle{} galaxies with highly eccentric orbits
could come from single mergers occurring between lookback times 8 and 9 Gyr.

This time coincides with the simulated \GE{} merger and also with the start of
the most significant growth of the thin disk. Interestingly, in our simulation
\GE{} causes an increase in star formation in the MW and also contributes a
significant amount of gas ($\sim2.9\times10^{10}$ \MsunInline{}) to the thin
disk. This scenario resembles that proposed for the two-infall model
\citep{spitoni_galactic_2019}.

Our findings show clearly that if a galaxy with global properties that are
similar to the MW is also selected to have the observationally expected assembly
history, it can provide insight into the early formation of the MW components.
Nonetheless, we expect that a higher resolution, zoom-in simulation of this
system will allow for a more detailed exploration of these processes and others,
such as those related with the bulge/bar formation and the effects of the dark
matter distribution.

\section*{Acknowledgements}
This project has received funding from the European Union’s Horizon 2020
Research and Innovation Programme under the Marie Sklodowska-Curie grant
agreement No. 734374 (LACEGAL). We acknowledge use of the Ragnar Cluster of
the Laboratory for Numerical Astrophysics of Universidad Andres Bello. A.H.
acknowledges financial support from NWO through a Vici personal grant. L.B.
acknowledges support from CONICYT FONDECYT/POSTDOCTORADO/3180359. P.B.T.
acknowledges partial support from UNAB personal grant.




\bibliographystyle{aasjournal}
\bibliography{gaia_eagle} 

\begin{thebibliography}{}
\expandafter\ifx\csname natexlab\endcsname\relax\def\natexlab#1{#1}\fi
\providecommand{\url}[1]{\href{#1}{#1}}
\providecommand{\dodoi}[1]{doi:~\href{http://doi.org/#1}{\nolinkurl{#1}}}
\providecommand{\doeprint}[1]{\href{http://ascl.net/#1}{\nolinkurl{http://ascl.net/#1}}}
\providecommand{\doarXiv}[1]{\href{https://arxiv.org/abs/#1}{\nolinkurl{https://arxiv.org/abs/#1}}}

\bibitem[{Abolfathi {et~al.}(2018)Abolfathi, Aguado, Aguilar, Allende~Prieto,
  Almeida, Ananna, Anders, Anderson, Andrews, Anguiano, {Arag{\'o}n-Salamanca},
  {Argudo-Fern{\'a}ndez}, Armengaud, Ata, Aubourg, {Avila-Reese}, Badenes,
  Bailey, Balland, Barger, {Barrera-Ballesteros}, Bartosz, Bastien, Bates,
  Baumgarten, Bautista, Beaton, Beers, Belfiore, Bender, Bernardi, Bershady,
  Beutler, Bird, Bizyaev, Blanc, Blanton, Blomqvist, Bolton, Boquien,
  Borissova, Bovy, Bradna~Diaz, Brandt, Brinkmann, Brownstein, Bundy,
  Burgasser, Burtin, Busca, Ca{\~n}as, {Cano-D{\'i}az}, Cappellari, Carrera,
  Casey, Cervantes~Sodi, Chen, Cherinka, Chiappini, Choi, Chojnowski, Chuang,
  Chung, Clerc, Cohen, Comerford, Comparat, {Correa do Nascimento}, {da Costa},
  Cousinou, Covey, Crane, {Cruz-Gonzalez}, Cunha, {da Silva Ilha}, Damke,
  Darling, Davidson, Dawson, {de Icaza Lizaola}, {de la Macorra}, {de la
  Torre}, De~Lee, {de Sainte Agathe}, Deconto~Machado, Dell'Agli, Delubac,
  {Diamond-Stanic}, Donor, Downes, Drory, {du Mas des Bourboux}, Duckworth,
  Dwelly, Dyer, Ebelke, Davis~Eigenbrot, Eisenstein, Elsworth, Emsellem,
  Eracleous, Erfanianfar, Escoffier, Fan, Fern{\'a}ndez~Alvar,
  {Fernandez-Trincado}, Fernando~Cirolini, Feuillet, Finoguenov, Fleming,
  {Font-Ribera}, Freischlad, Frinchaboy, Fu, G{\'o}mez Maqueo~Chew, Galbany,
  Garc{\'i}a~P{\'e}rez, {Garcia-Dias}, {Garc{\'i}a-Hern{\'a}ndez},
  Garma~Oehmichen, Gaulme, Gelfand, {Gil-Mar{\'i}n}, Gillespie, Goddard,
  Gonz{\'a}lez~Hern{\'a}ndez, {Gonzalez-Perez}, Grabowski, Green, Grier,
  Gueguen, Guo, Guy, Hagen, Hall, Harding, Hasselquist, Hawley, Hayes, Hearty,
  Hekker, Hernandez, Hernandez~Toledo, Hogg, {Holley-Bockelmann}, Holtzman,
  Hou, Hsieh, Hunt, Hutchinson, Hwang, Jimenez~Angel, Johnson, Jones,
  J{\"o}nsson, Jullo, Khan, Kinemuchi, Kirkby, Kirkpatrick, Kitaura, Knapp,
  Kneib, Kollmeier, Lacerna, Lane, Lang, Law, Le~Goff, Lee, Li, Li, Lian,
  Liang, Lima, Lin, Long, Lucatello, Lundgren, Mackereth, MacLeod, Mahadevan,
  Maia, Majewski, Manchado, Maraston, Mariappan, {Marques-Chaves}, Masseron,
  Masters, McDermid, McGreer, Melendez, {Meneses-Goytia}, Merloni, Merrifield,
  Meszaros, Meza, Minchev, Minniti, Mueller, {Muller-Sanchez}, Muna, Mu{\~n}oz,
  Myers, Nair, Nandra, Ness, Newman, Nichol, Nidever, Nitschelm, Noterdaeme,
  O'Connell, Oelkers, Oravetz, Oravetz, Ort{\'i}z, Osorio, Pace, Padilla,
  {Palanque-Delabrouille}, Palicio, Pan, Pan, Parikh, P{\^a}ris, Park, Peirani,
  {Pellejero-Ibanez}, Penny, Percival, {Perez-Fournon}, Petitjean, Pieri,
  Pinsonneault, Pisani, Prada, Prakash, Queiroz, Raddick, Raichoor,
  Barboza~Rembold, Richstein, Riffel, Riffel, Rix, Robin, Rodr{\'i}guez~Torres,
  {Rom{\'a}n-Z{\'u}{\~n}iga}, Ross, Rossi, Ruan, Ruggeri, Ruiz, Salvato,
  S{\'a}nchez, S{\'a}nchez, Sanchez~Almeida, {S{\'a}nchez-Gallego},
  Santana~Rojas, Santiago, Schiavon, Schimoia, Schlafly, Schlegel, Schneider,
  Schuster, Schwope, Seo, Serenelli, Shen, Shen, Shetrone, Shull,
  Silva~Aguirre, Simon, Skrutskie, Slosar, Smethurst, Smith, Sobeck, Somers,
  Souter, Souto, Spindler, Stark, Stassun, Steinmetz, Stello,
  {Storchi-Bergmann}, Streblyanska, Stringfellow, Su{\'a}rez, Sun, Szigeti,
  {Taghizadeh-Popp}, Talbot, Tang, Tao, Tayar, Tembe, Teske, Thakar, Thomas,
  Tissera, Tojeiro, Tremonti, Troup, Urry, Valenzuela, {van den Bosch},
  {Vargas-Gonz{\'a}lez}, {Vargas-Maga{\~n}a}, Vazquez, Villanova, Vogt, Wake,
  Wang, Weaver, Weijmans, Weinberg, Westfall, Whelan, Wilcots, Wild, Williams,
  Wilson, {Wood-Vasey}, Wylezalek, Xiao, Yan, Yang, Ybarra, Y{\`e}che,
  Zakamska, Zamora, Zarrouk, Zasowski, Zhang, Zhao, Zhao, Zheng, Zheng, Zhou,
  Zhu, Zinn, \& Zou}]{abolfathi_fourteenth_2018}
Abolfathi, B., Aguado, D.~S., Aguilar, G., {et~al.} 2018, ApJS, 235, 42,
  \dodoi{10.3847/1538-4365/aa9e8a}

\bibitem[{Barnes(2004)}]{barnes_shock-induced_2004}
Barnes, J.~E. 2004, MNRAS, 350, 798, \dodoi{10.1111/j.1365-2966.2004.07725.x}

\bibitem[{Belokurov {et~al.}(2018)Belokurov, Erkal, Evans, Koposov, \&
  Deason}]{belokurov_co-formation_2018}
Belokurov, V., Erkal, D., Evans, N.~W., Koposov, S.~E., \& Deason, A.~J. 2018,
  MNRAS, 478, 611, \dodoi{10.1093/mnras/sty982}

\bibitem[{Buder {et~al.}(2018)Buder, Asplund, Duong, Kos, Lind, Ness, Sharma,
  {Bland-Hawthorn}, Casey, {de Silva}, D'Orazi, Freeman, Lewis, Lin, Martell,
  Schlesinger, Simpson, Zucker, Zwitter, Amarsi, Anguiano, Carollo, Casagrande,
  {\v C}otar, Cottrell, {da Costa}, Gao, Hayden, Horner, Ireland, Kafle,
  Munari, Nataf, Nordlander, Stello, Ting, Traven, Watson, Wittenmyer, Wyse,
  Yong, Zinn, {\v Z}erjal, \& Collaboration}]{buder_galah_2018}
Buder, S., Asplund, M., Duong, L., {et~al.} 2018, MNRAS, 478, 4513,
  \dodoi{10.1093/mnras/sty1281}

\bibitem[{Carollo {et~al.}(2018)Carollo, Tissera, Beers, Gudin, Gibson,
  Freeman, \& Monachesi}]{carollo_origin_2018}
Carollo, D., Tissera, P.~B., Beers, T.~C., {et~al.} 2018, ApJ, 859, L7,
  \dodoi{10.3847/2041-8213/aac2dc}

\bibitem[{Carollo {et~al.}(2019)Carollo, Chiba, Ishigaki, Freeman, Beers, Lee,
  Tissera, Battistini, \& Primas}]{carollo_evidence_2019}
Carollo, D., Chiba, M., Ishigaki, M., {et~al.} 2019, arXiv e-prints,
  arXiv:1904.04881

\bibitem[{Crain {et~al.}(2015)Crain, Schaye, Bower, Furlong, Schaller, Theuns,
  Dalla~Vecchia, Frenk, McCarthy, Helly, Jenkins, {Rosas-Guevara}, White, \&
  Trayford}]{crain_eagle_2015}
Crain, R.~A., Schaye, J., Bower, R.~G., {et~al.} 2015, MNRAS, 450, 1937,
  \dodoi{10.1093/mnras/stv725}

\bibitem[{Cui {et~al.}(2012)Cui, Zhao, Chu, Li, Li, Zhang, Su, Yao, Wang, Xing,
  Li, Zhu, Wang, Gu, Luo, Xu, Zhang, Liu, Zhang, Yang, Cao, Chen, Chen, Chen,
  Chen, Chu, Feng, Gong, Hou, Hu, Hu, Hu, Jia, Jiang, Jiang, Jiang, Jin, Li,
  Li, Li, Liu, Liu, Lu, Mao, Men, Qi, Qi, Shi, Tang, Tao, Wang, Wang, Wang,
  Wang, Wang, Wang, Wang, Wang, Wang, Wang, Wang, Wang, Xu, Xu, Yang, Yu, Yuan,
  Yuan, Zhai, Zhang, Zhang, Zhang, Zhao, Zhou, Zhou, Zhu, \&
  Zou}]{cui_large_2012}
Cui, X.-Q., Zhao, Y.-H., Chu, Y.-Q., {et~al.} 2012, Research in Astronomy and
  Astrophysics, 12, 1197, \dodoi{10.1088/1674-4527/12/9/003}

\bibitem[{Fattahi {et~al.}(2019)Fattahi, Belokurov, Deason, Frenk, G{\'o}mez,
  Grand, Marinacci, Pakmor, \& Springel}]{fattahi_origin_2019}
Fattahi, A., Belokurov, V., Deason, A.~J., {et~al.} 2019, MNRAS, 484, 4471,
  \dodoi{10.1093/mnras/stz159}

\bibitem[{{Fern{\'a}ndez-Alvar} {et~al.}(2019){Fern{\'a}ndez-Alvar}, Tissera,
  Carigi, Schuster, Beers, \& Belokurov}]{fernandez-alvar_assembly_2019}
{Fern{\'a}ndez-Alvar}, E., Tissera, P.~B., Carigi, L., {et~al.} 2019, MNRAS,
  485, 1745, \dodoi{10.1093/mnras/stz443}

\bibitem[{{Gaia Collaboration} {et~al.}(2018){Gaia Collaboration}, Brown,
  Vallenari, Prusti, {de Bruijne}, Babusiaux, {Bailer-Jones}, Biermann, Evans,
  Eyer, Jansen, Jordi, Klioner, Lammers, Lindegren, Luri, Mignard, Panem,
  Pourbaix, Randich, Sartoretti, Siddiqui, Soubiran, {van Leeuwen}, Walton,
  Arenou, Bastian, Cropper, Drimmel, Katz, Lattanzi, Bakker, Cacciari,
  Casta{\~n}eda, Chaoul, Cheek, De~Angeli, Fabricius, Guerra, Holl, Masana,
  Messineo, Mowlavi, Nienartowicz, Panuzzo, Portell, Riello, Seabroke, Tanga,
  Th{\'e}venin, {Gracia-Abril}, Comoretto, {Garcia-Reinaldos}, Teyssier,
  Altmann, Andrae, Audard, {Bellas-Velidis}, Benson, Berthier, Blomme, Burgess,
  Busso, Carry, Cellino, Clementini, Clotet, Creevey, Davidson, De~Ridder,
  Delchambre, Dell'Oro, Ducourant, {Fern{\'a}ndez-Hern{\'a}ndez}, Fouesneau,
  Fr{\'e}mat, Galluccio, {Garc{\'i}a-Torres}, {Gonz{\'a}lez-N{\'u}{\~n}ez},
  {Gonz{\'a}lez-Vidal}, Gosset, Guy, Halbwachs, Hambly, Harrison,
  Hern{\'a}ndez, Hestroffer, Hodgkin, Hutton, Jasniewicz,
  {Jean-Antoine-Piccolo}, Jordan, Korn, {Krone-Martins}, Lanzafame, Lebzelter,
  L{\"o}ffler, Manteiga, Marrese, {Mart{\'i}n-Fleitas}, Moitinho, Mora,
  Muinonen, Osinde, Pancino, Pauwels, Petit, {Recio-Blanco}, Richards,
  Rimoldini, Robin, Sarro, Siopis, Smith, Sozzetti, S{\"u}veges, Torra, {van
  Reeven}, Abbas, Abreu~Aramburu, Accart, Aerts, Altavilla, {\'A}lvarez,
  Alvarez, Alves, Anderson, Andrei, Anglada~Varela, Antiche, Antoja, Arcay,
  Astraatmadja, Bach, Baker, {Balaguer-N{\'u}{\~n}ez}, Balm, Barache, Barata,
  Barbato, Barblan, Barklem, Barrado, Barros, Barstow,
  Bartholom{\'e}~Mu{\~n}oz, Bassilana, Becciani, Bellazzini, Berihuete,
  Bertone, Bianchi, Bienaym{\'e}, {Blanco-Cuaresma}, Boch, Boeche, Bombrun,
  Borrachero, Bossini, Bouquillon, Bourda, Bragaglia, Bramante, Breddels,
  Bressan, Brouillet, Br{\"u}semeister, Brugaletta, Bucciarelli, Burlacu,
  Busonero, Butkevich, Buzzi, Caffau, Cancelliere, Cannizzaro, {Cantat-Gaudin},
  Carballo, Carlucci, Carrasco, Casamiquela, Castellani, {Castro-Ginard},
  Charlot, Chemin, Chiavassa, Cocozza, Costigan, Cowell, Crifo, Crosta,
  Crowley, Cuypers, Dafonte, Damerdji, Dapergolas, David, David, {de Laverny},
  De~Luise, De~March, {de Martino}, {de Souza}, {de Torres}, Debosscher, {del
  Pozo}, Delbo, Delgado, Delgado, Di~Matteo, Diakite, Diener, Distefano,
  Dolding, Drazinos, Dur{\'a}n, Edvardsson, Enke, Eriksson, Esquej,
  Eynard~Bontemps, Fabre, Fabrizio, Faigler, Falc{\~a}o, Farr{\`a}s~Casas,
  Federici, Fedorets, Fernique, Figueras, Filippi, Findeisen, Fonti, Fraile,
  Fraser, Fr{\'e}zouls, Gai, Galleti, Garabato, {Garc{\'i}a-Sedano}, Garofalo,
  Garralda, Gavel, Gavras, Gerssen, Geyer, Giacobbe, Gilmore, Girona,
  Giuffrida, Glass, Gomes, Granvik, Gueguen, Guerrier, Guiraud,
  {Guti{\'e}rrez-S{\'a}nchez}, Haigron, Hatzidimitriou, Hauser, Haywood,
  Heiter, Helmi, Heu, Hilger, Hobbs, Hofmann, Holland, Huckle, Hypki, Icardi,
  Jan{\ss}en, {Jevardat de Fombelle}, Jonker, Juh{\'a}sz, Julbe, Karampelas,
  Kewley, Klar, Kochoska, Kohley, Kolenberg, Kontizas, Kontizas, Koposov,
  Kordopatis, {Kostrzewa-Rutkowska}, Koubsky, Lambert, Lanza, Lasne, Lavigne,
  Le~Fustec, {Le Poncin-Lafitte}, Lebreton, Leccia, Leclerc, {Lecoeur-Taibi},
  Lenhardt, Leroux, Liao, Licata, Lindstr{\o}m, Lister, Livanou, Lobel,
  L{\'o}pez, Managau, Mann, Mantelet, Marchal, Marchant, Marconi, Marinoni,
  Marschalk{\'o}, Marshall, Martino, Marton, Mary, Massari, Matijevi{\v c},
  Mazeh, McMillan, Messina, Michalik, Millar, Molina, Molinaro, Moln{\'a}r,
  Montegriffo, Mor, Morbidelli, Morel, Morris, Mulone, Muraveva, Musella,
  Nelemans, Nicastro, Noval, O'Mullane, Ord{\'e}novic,
  {Ord{\'o}{\~n}ez-Blanco}, Osborne, Pagani, Pagano, Pailler, Palacin,
  Palaversa, Panahi, Pawlak, Piersimoni, Pineau, Plachy, Plum, Poggio,
  Poujoulet, Pr{\v s}a, Pulone, Racero, Ragaini, Rambaux, {Ramos-Lerate},
  Regibo, Reyl{\'e}, Riclet, Ripepi, Riva, Rivard, Rixon, Roegiers, Roelens,
  {Romero-G{\'o}mez}, Rowell, Royer, {Ruiz-Dern}, Sadowski,
  Sagrist{\`a}~Sell{\'e}s, Sahlmann, Salgado, Salguero, Sanna, {Santana-Ros},
  Sarasso, Savietto, Schultheis, Sciacca, Segol, Segovia, S{\'e}gransan, Shih,
  Siltala, Silva, Smart, Smith, Solano, Solitro, Sordo, Soria~Nieto, Souchay,
  Spagna, Spoto, Stampa, Steele, Steidelm{\"u}ller, Stephenson, Stoev, Suess,
  Surdej, Szabados, {Szegedi-Elek}, Tapiador, Taris, Tauran, Taylor, Teixeira,
  Terrett, Teyssandier, Thuillot, Titarenko, Torra~Clotet, Turon, Ulla,
  Utrilla, Uzzi, Vaillant, Valentini, Valette, {van Elteren}, Van~Hemelryck,
  {van Leeuwen}, Vaschetto, Vecchiato, Veljanoski, Viala, Vicente, Vogt, {von
  Essen}, Voss, Votruba, Voutsinas, Walmsley, Weiler, Wertz, Wevers,
  Wyrzykowski, Yoldas, {\v Z}erjal, Ziaeepour, Zorec, Zschocke, Zucker,
  Zurbach, \& Zwitter}]{collaboration_gaia_2018}
{Gaia Collaboration}, Brown, A. G.~A., Vallenari, A., {et~al.} 2018, A\&A, 616,
  A1, \dodoi{10.1051/0004-6361/201833051}

\bibitem[{Gallart {et~al.}(2019)Gallart, Bernard, Brook, {Ruiz-Lara}, Cassisi,
  Hill, \& Monelli}]{gallart_uncovering_2019}
Gallart, C., Bernard, E.~J., Brook, C.~B., {et~al.} 2019, Nat Astron, 407,
  \dodoi{10.1038/s41550-019-0829-5}

\bibitem[{G{\'o}mez {et~al.}(2017)G{\'o}mez, Grand, Monachesi, White,
  Bustamante, Marinacci, Pakmor, Simpson, Springel, \&
  Frenk}]{gomez_lessons_2017}
G{\'o}mez, F.~A., Grand, R. J.~J., Monachesi, A., {et~al.} 2017, MNRAS, 472,
  3722, \dodoi{10.1093/mnras/stx2149}

\bibitem[{Hawkins {et~al.}(2014)Hawkins, Jofr{\'e}, Gilmore, \&
  Masseron}]{hawkins_relative_2014}
Hawkins, K., Jofr{\'e}, P., Gilmore, G., \& Masseron, T. 2014, MNRAS, 445,
  2575, \dodoi{10.1093/mnras/stu1910}

\bibitem[{Haywood {et~al.}(2018)Haywood, Di~Matteo, Lehnert, Snaith,
  Khoperskov, \& G{\'o}mez}]{haywood_disguise_2018}
Haywood, M., Di~Matteo, P., Lehnert, M.~D., {et~al.} 2018, ApJ, 863, 113,
  \dodoi{10.3847/1538-4357/aad235}

\bibitem[{Helmi {et~al.}(2018)Helmi, Babusiaux, Koppelman, Massari, Veljanoski,
  \& Brown}]{helmi_merger_2018}
Helmi, A., Babusiaux, C., Koppelman, H.~H., {et~al.} 2018, Nature, 563, 85,
  \dodoi{10.1038/s41586-018-0625-x}

\bibitem[{{Herzog-Arbeitman} {et~al.}(2018){Herzog-Arbeitman}, Lisanti, Madau,
  \& Necib}]{herzog-arbeitman_empirical_2018}
{Herzog-Arbeitman}, J., Lisanti, M., Madau, P., \& Necib, L. 2018, Phys. Rev.
  Lett., 120, 041102, \dodoi{10.1103/PhysRevLett.120.041102}

\bibitem[{Ibata {et~al.}(1994)Ibata, Gilmore, \& Irwin}]{ibata_dwarf_1994}
Ibata, R.~A., Gilmore, G., \& Irwin, M.~J. 1994, Nature, 370, 194,
  \dodoi{10.1038/370194a0}

\bibitem[{Katz {et~al.}(1994)Katz, Quinn, Bertschinger, \&
  Gelb}]{katz_formation_1994}
Katz, N., Quinn, T., Bertschinger, E., \& Gelb, J.~M. 1994, MNRAS, 270, L71,
  \dodoi{10.1093/mnras/270.1.L71}

\bibitem[{Koppelman {et~al.}(2018)Koppelman, Helmi, \&
  Veljanoski}]{koppelman_one_2018}
Koppelman, H., Helmi, A., \& Veljanoski, J. 2018, ApJ, 860, L11,
  \dodoi{10.3847/2041-8213/aac882}

\bibitem[{Kunder {et~al.}(2017)Kunder, Kordopatis, Steinmetz, Zwitter,
  McMillan, Casagrande, Enke, Wojno, Valentini, Chiappini, Matijevi{\v c},
  Siviero, {de Laverny}, {Recio-Blanco}, Bijaoui, Wyse, Binney, Grebel, Helmi,
  Jofre, Antoja, Gilmore, Siebert, Famaey, Bienaym{\'e}, Gibson, Freeman,
  Navarro, Munari, Seabroke, Anguiano, {\v Z}erjal, Minchev, Reid,
  {Bland-Hawthorn}, Kos, Sharma, Watson, Parker, Scholz, Burton, Cass, Hartley,
  Fiegert, Stupar, Ritter, Hawkins, Gerhard, Chaplin, Davies, Elsworth, Lund,
  Miglio, \& Mosser}]{kunder_radial_2017}
Kunder, A., Kordopatis, G., Steinmetz, M., {et~al.} 2017, AJ, 153, 75,
  \dodoi{10.3847/1538-3881/153/2/75}

\bibitem[{Lambas {et~al.}(2003)Lambas, Tissera, Alonso, \&
  Coldwell}]{lambas_galaxy_2003-1}
Lambas, D.~G., Tissera, P.~B., Alonso, M.~S., \& Coldwell, G. 2003, MNRAS, 346,
  1189, \dodoi{10.1111/j.1365-2966.2003.07179.x}

\bibitem[{Larson \& Tinsley(1978)}]{larson_star_1978}
Larson, R.~B., \& Tinsley, B.~M. 1978, Astrophys. J., 219, 46,
  \dodoi{10.1086/155753}

\bibitem[{Mackereth {et~al.}(2019)Mackereth, Schiavon, Pfeffer, Hayes, Bovy,
  Anguiano, Allende~Prieto, Hasselquist, Holtzman, Johnson, Majewski,
  O'Connell, Shetrone, Tissera, \&
  {Fern{\'a}ndez-Trincado}}]{mackereth_origin_2019}
Mackereth, J.~T., Schiavon, R.~P., Pfeffer, J., {et~al.} 2019, MNRAS, 482,
  3426, \dodoi{10.1093/mnras/sty2955}

\bibitem[{McAlpine {et~al.}(2016)McAlpine, Helly, Schaller, Trayford, Qu,
  Furlong, Bower, Crain, Schaye, Theuns, Vecchia, Frenk, McCarthy, Jenkins,
  {Rosas-Guevara}, White, Baes, Camps, \& Lemson}]{mcalpine_eagle_2016}
McAlpine, S., Helly, J.~C., Schaller, M., {et~al.} 2016, A\&C, 15, 72,
  \dodoi{10.1016/j.ascom.2016.02.004}

\bibitem[{Mihos \& Hernquist(1996)}]{mihos_gasdynamics_1996-1}
Mihos, J.~C., \& Hernquist, L. 1996, The Astrophysical Journal, 464, 641,
  \dodoi{10.1086/177353}

\bibitem[{Myeong {et~al.}(2018)Myeong, Evans, Belokurov, Sanders, \&
  Koposov}]{myeong_milky_2018}
Myeong, G.~C., Evans, N.~W., Belokurov, V., Sanders, J.~L., \& Koposov, S.~E.
  2018, ApJ, 856, L26, \dodoi{10.3847/2041-8213/aab613}

\bibitem[{Navarro \& White(1994)}]{navarro_simulations_1994}
Navarro, J.~F., \& White, S. D.~M. 1994, MNRAS, 267, 401,
  \dodoi{10.1093/mnras/267.2.401}

\bibitem[{Necib {et~al.}(2018)Necib, Lisanti, {Garrison-Kimmel}, Wetzel,
  Sanderson, Hopkins, {Faucher-Gigu{\`e}re}, \& Kere{\v s}}]{necib_under_2018}
Necib, L., Lisanti, M., {Garrison-Kimmel}, S., {et~al.} 2018, arXiv e-prints,
  arXiv:1810.12301

\bibitem[{Posti \& Helmi(2019)}]{posti_mass_2019}
Posti, L., \& Helmi, A. 2019, A\&A, 621, A56,
  \dodoi{10.1051/0004-6361/201833355}

\bibitem[{Schaye {et~al.}(2015)Schaye, Crain, Bower, Furlong, Schaller, Theuns,
  Dalla~Vecchia, Frenk, McCarthy, Helly, Jenkins, {Rosas-Guevara}, White, Baes,
  Booth, Camps, Navarro, Qu, Rahmati, Sawala, Thomas, \&
  Trayford}]{schaye_eagle_2015}
Schaye, J., Crain, R.~A., Bower, R.~G., {et~al.} 2015, MNRAS, 446, 521,
  \dodoi{10.1093/mnras/stu2058}

\bibitem[{Sillero {et~al.}(2017)Sillero, Tissera, Lambas, \&
  {Michel-Dansac}}]{sillero_evolution_2017}
Sillero, E., Tissera, P.~B., Lambas, D.~G., \& {Michel-Dansac}, L. 2017, MNRAS,
  472, 4404, \dodoi{10.1093/mnras/stx2265}

\bibitem[{Spitoni {et~al.}(2019)Spitoni, Silva~Aguirre, Matteucci, Calura, \&
  Grisoni}]{spitoni_galactic_2019}
Spitoni, E., Silva~Aguirre, V., Matteucci, F., Calura, F., \& Grisoni, V. 2019,
  A\&A, 623, A60, \dodoi{10.1051/0004-6361/201834188}

\bibitem[{Tissera {et~al.}(2012)Tissera, White, \&
  Scannapieco}]{tissera_chemical_2012}
Tissera, P.~B., White, S. D.~M., \& Scannapieco, C. 2012, MNRAS, 420, 255,
  \dodoi{10.1111/j.1365-2966.2011.20028.x}

\bibitem[{Villalobos \& Helmi(2008)}]{villalobos_simulations_2008}
Villalobos, {\'A}., \& Helmi, A. 2008, MNRAS, 391, 1806,
  \dodoi{10.1111/j.1365-2966.2008.13979.x}

\bibitem[{Vincenzo {et~al.}(2019)Vincenzo, Spitoni, Calura, Matteucci,
  Silva~Aguirre, Miglio, \& Cescutti}]{vincenzo_fall_2019}
Vincenzo, F., Spitoni, E., Calura, F., {et~al.} 2019, MNRAS, 487, L47,
  \dodoi{10.1093/mnrasl/slz070}

\bibitem[{Watkins {et~al.}(2019)Watkins, {van der Marel}, Sohn, \&
  Evans}]{watkins_evidence_2019}
Watkins, L.~L., {van der Marel}, R.~P., Sohn, S.~T., \& Evans, N.~W. 2019, ApJ,
  873, 118, \dodoi{10.3847/1538-4357/ab089f}

\bibitem[{Zhao {et~al.}(2012)Zhao, Zhao, Chu, Jing, \& Deng}]{zhao_lamost_2012}
Zhao, G., Zhao, Y.-H., Chu, Y.-Q., Jing, Y.-P., \& Deng, L.-C. 2012, Research
  in Astronomy and Astrophysics, 12, 723, \dodoi{10.1088/1674-4527/12/7/002}

\end{thebibliography}






\end{document}